\documentclass[12pt]{article}
\usepackage{amssymb,graphicx,amsmath,array,verbatim,arydshln}
\usepackage[dvipdfmx, bookmarks=false]{hyperref}

\newcommand{\beq}{\begin{eqnarray}}
\newcommand{\eeq}{\end{eqnarray}}
\newcommand{\p}{\partial}
\newcommand{\NF}{N_{\rm F}}

\newcommand{\vs}[1]{\vspace{#1 mm}}
\newcommand{\hs}[1]{\hspace{#1 mm}}
\newcommand{\bpm}{\begin{pmatrix}}
\newcommand{\epm}{\end{pmatrix}}
\newcommand{\Z}{\mathbb{Z}}

\newcommand{\C}{\mathbb{C}}

\newcommand{\D}{\mathcal D}

\newcommand{\ba}{\left(\begin{array}}
\newcommand{\ea}{\end{array} \right)}

\makeatletter
\@addtoreset{equation}{section}

\makeatother

\usepackage[usenames]{color}

\begin{document}
\title{Massive Nambu-Goldstone Fermions and Bosons \\
for Non-relativistic Superconformal Symmetry: \\
Jackiw-Pi Vortices in a Trap}

\author{Toshiaki Fujimori, Muneto Nitta, Keisuke Ohashi 
\\{\it\small 
Department of Physics, and Research and Education Center for Natural Sciences,} 
\\{\it\small 
Keio University, Hiyoshi 4-1-1, Yokohama, Kanagawa 223-8521, Japan}}

\maketitle

\begin{abstract} \vs{10}
We discuss a supersymmetric extension of 
a non-relativistic Chern-Simons matter theory, 
known as the SUSY Jackiw-Pi model, 
in a harmonic trap. 
We show that 
the non-relativistic version of the superconformal symmetry, 
called the super-Schr\"odinger symmetry, 
is not spoiled by an external field 
including the harmonic potential.
It survives as a modified symmetry 
whose generators have explicit time dependences
determined by the strength of the trap, 
the rotation velocity of the system 
and the fermion number chemical potential. 
We construct 1/3 BPS states of trapped Jackiw-Pi vortices 
preserving a part of the modified superconformal symmetry 
and discuss fluctuations around static BPS configurations. 
In addition to the bosonic massive Nambu-Goldstone modes, 
we find that there exist massive Nambu-Goldstone fermions 
associated with broken generators of 
the modified super-Schr\"odinger symmetry. 
Furthermore, we find that eigenmodes form 
supermultiplets of a modified supersymmetry
preserved by the static BPS backgrounds. 
As a consequence of the modified supersymmetry, 
infinite towers of explicit spectra can be found 
for eigenmodes corresponding to 
bosonic and fermionic lowest Landau levels. 

\end{abstract}

\newpage

\tableofcontents


\section{Introduction} \label{sec:Intro}
Non-trivial external background fields are 
useful tools to study various aspects of field theories. 
When a generic background field is turned on in a physical system, 
it may break a symmetry of the system and 
drastically change the structure of the model.
However it has been known that 
if an external field can be viewed as a chemical potential term 
associated with a conserved charge, 
a version of the Nambu-Goldstone (NG) theorem can 
still be applied even when a symmetry appears
explicitly broken by the external field. 
The crucial difference from the standard NG theorem 
is that the corresponding NG mode in this case
has a non-vanishing mass 
precisely determined by the symmetry algebra. 
Such \textit{massive Nambu-Goldstone bosons}
have been discussed in \cite{Nicolis:2012vf, Nicolis:2013sga, Watanabe:2013uya, Takahashi:2014vua} and 
scattering amplitudes of massive NG modes 
was recently studied in Ref.\,\cite{Brauner:2017gkr}.

In Refs.\,\cite{Pitaevskii:1997, Ghosh:2001an, Ripoll:2001, Ohashi:2017vcy, Takahashi:2017ruq}, 
various properties of the massive NG modes 
associated with the non-relativistic conformal symmetry, 
called the Schr\"odinger symmetry \cite{Hagen:1972pd, Niederer:1972zz} , 
have been revealed in the $2+1$ dimensional 
non-linear Schr\"odinger system in a harmonic trap. 
One of the most important observations is that 
the Schr\"odinger symmetry survives
even in the presence of external background fields 
including the harmonic potential. 
More precisely, a modified Schr\"odinger symmetry 
generated by time dependent operators 
remains in such a background. 
In general, when a symmetry generated by an operator 
with an explicit time dependence is spontaneously broken, 
the associated NG modes has a non-vanishing mass 
determined by the commutation relation between
the corresponding broken generator and the Hamiltonian. 
As in the case of the Lorentz and Galilean symmetry, 
a time dependent symmetry can be used 
to study dynamical properties of the system. 
For example, 
in the non-linear Schr\"odinger system in a harmonic trap, 
time dependent solutions can be generated from static ones 
by applying the time dependent modified Schr\"odinger symmetry. 

In this paper, we discuss 
the supersymmetric Jackiw-Pi model 
and study vortices and massive NG modes 
in a non-trivial background. 
The Jackiw-Pi model is a field theoretic framework 
describing anyons in terms of 
the non-linear Schr\"odinger system
coupled with a Chern-Simons gauge field \cite{Jackiw:1990mb}. 
As with the standard non-linear Schr\"odinger model, 
the Jackiw-Pi model has 
a modified (time dependent) Schr\"odinger symmetry 
in various backgrounds. 
Non-topological vortex solutions,  
called the Jackiw-Pi vortices \cite{Jackiw:1990tz, Dunne:1990qe},  
have been discussed in such backgrounds
\cite{Ezawa:1991sh, Ezawa:1991qn, Jackiw:1991hh, Jackiw:1991ns, Ezawa:1991xf, Duval:1994vh, Doroud:2015fsz},
and in particular time dependent solutions were constructed 
by making use of maps between the models 
with and without the external fields. 

The Jackiw-Pi model without background field 
has a supersymmetric extension
which possesses a non-relativistic superconformal symmetry, 
called the super-Schr\"odinger symmetry 
\cite{Gauntlett:1990xq, Leblanc:1992wu, Duval:1993hs}. 
In this paper, we show that external background fields 
corresponding to the harmonic potential, the spatial rotation, 
the flavor and fermion number chemical potentials,
do not spoil the superconformal symmetry as well as 
the Schr\"odinger symmetry. 
In the presence of such external fields, 
the whole super-Schr\"odinger symmetry 
becomes a time dependent symmetry of the type
which has been discussed 
in the context of the supersymmetric harmonic oscillator 
in quantum mechanics \cite{Beckers:1986ty, Beckers:1987xr}. 
  
We also discuss the Jackiw-Pi vortices 
in the non-trivial background fields and 
construct their 1/3 BPS states  
which is invariant under a part of 
the time dependent supersymmetry. 
The moduli matrix formalism, 
which has been used to describe 
the moduli space of non-Abelian vortices 
\cite{Eto:2005yh, Eto:2006pg, Isozumi:2004vg, Eto:2004rz, Eto:2006cx, Eto:2006db}, 
can also be applied to write down 
a formal solution of the 1/3 BPS equation in this system. 
For each choice of a holomorphic matrix $H_0(z)$, 
we can obtain a BPS configuration of 
trapped Jackiw-Pi vortices 
by solving the Gauss law equation. 
Generic 1/3 BPS solutions turn out to be 
Q-soliton-like configurations, that is, 
they are time dependent stationary configurations
stabilized by conserved charges. 
They are new time dependent solutions 
which are different from the known solutions
obtained by using the maps between the models 
with and without the external fields \cite{Ezawa:1991sh, Ezawa:1991qn, Jackiw:1991hh, Jackiw:1991ns, Ezawa:1991xf, Duval:1994vh}.  

The BPS solutions becomes static configurations 
if $H_0(z)$ takes one of special forms 
corresponding to the fixed points of the spacial and flavor rotation. 
We discuss fluctuations around them  
and show that bosonic and fermionic eigenmodes
form supermultiplets of the unbroken time dependent supersymmetry. 
There are two types of supermultiplets: 
one is a generic supermultiplet composed of 
a pair of bosonic and fermionic modes and
the other is a short supermultiplet consisting 
only of a bosonic component. 
In particular, we show that 
in addition to bosonic massive NG modes 
associated with spontaneously broken generators of 
the modified Schr\"odinger symmetry, 
there exist \textit{massive Goldstino} 
corresponding to spontaneously broken modified supercharges.
They consistently form supermultiplets 
as expected from the super-Schr\"odinger algebra.
In addition to those massive NG modes, 
we exactly derive eigenvalue spectra 
of infinite towers of short and long supermultiplets 
corresponding to the bosonic and fermionic lowest Landau levels,
respectively.  

The organization of the paper is as follows. 
In Sec.\,\ref{sec:trap}, 
we briefly review the super-Schr\"odinger symmetry 
in the supersymmetric Jackiw-Pi model 
and show that there exists 
a modified super-Schr\"odinger symmetry 
even in the presence of 
generalized chemical potential terms
including the harmonic potential. 
In Sec.\,\ref{sec:BPS}, 
we discuss 1/3 BPS solutions 
of trapped non-Abelian Jackiw-Pi vortices
which preserve a part of the modified superconformal symmetry. 
By applying the moduli matrix formalism,
we write down formal solutions and 
show that static configurations correspond 
to fixed points of the rotation and flavor symmetry. 
In Sec.\,\ref{sec:fluc}, 
we investigate fluctuations around static BPS backgrounds 
and elucidate the structure of supermultiplets of eigenmodes
including bosonic and fermionic massive NG modes. 
Sec.\,\ref{sec:summary} is devoted to a summary and discussions. 

\section{Supersymmetric Jackiw-Pi model in a harmonic trap}
\label{sec:trap}
\subsection{SUSY Jackiw-Pi model and super-Schr\"odinger symmetry}
The supersymmetric Jackiw-Pi model consists of 
a gauge field and pairs of bosonic and fermionic matter fields. 
For simplicity, we consider the case of 
a $U(N)$ gauge field $A_\mu$ 
with $\NF$ matter pairs $(\phi_I, \psi_I)~(I=1,\cdots,\NF)$ 
in the $(\mathbf N, \mathbf N_{\rm F})$ representation of 
the $U(N)$ gauge group and the $SU(\NF)$ flavor symmetry. 
It would be straightforward to extend 
the following discussion to more general settings.
By using $N$-by-$\NF$ matrix notation for the matter fields
\beq
\phi \equiv (\phi_1, \phi_2, \cdots, \phi_{\NF}), \hs{10}
\psi \equiv (\psi_1, \psi_2, \cdots, \psi_{\NF}), 
\eeq
the action of the supersymmetric Jackiw-Pi model can be written as
\beq
S = \int dt d^2 x \, {\rm Tr} 
\Bigg[ 
\phi^\dagger \hat \Delta_0 \phi 
+ \frac{1}{m} \psi^\dagger \hat \Delta_0 \psi 
- \frac{\pi}{km} M^2 
- \frac{\pi}{k m^2} \psi^\dagger \, Y \psi 
\Bigg] 
+ k S_{\rm CS}, 
\label{eq:JP_action}
\eeq
where the trace is taken over the flavor indices.  
Just for notational convenience, 
we have introduced 
the $\NF$-by-$\NF$ matrix $M$ and 
the $N$-by-$N$ matrix $Y$ deffined by
\beq
M \equiv \phi^\dagger \phi + \frac{1}{m} \psi^\dagger \psi, \hs{10}
Y \equiv \phi \phi^\dagger - \frac{1}{m} \psi \psi^\dagger + \frac{k}{\pi} i F_{z \bar z}. 
\eeq
The symbol $\hat \Delta_0$ denotes 
the differential operator which gives 
the standard non-relativistic kinetic term 
\beq
\hat \Delta_0 \equiv i \D_t + \frac{1}{m} \left( \D_z \D_{\bar z} + \D_{\bar z} \D_z \right). 
\eeq
The covariant derivative and the field strength are defined by 
$\D_\mu \phi \equiv (\p_\mu + i A_\mu) \phi$, 
$\D_\mu \phi^\dagger \equiv 
\p_\mu \phi^\dagger - i \phi^\dagger A_\mu$ and 
$F_{\mu \nu} \equiv - i [\D_\mu, \D_\nu]$, etc. 
The parameter $k$ is the Chern-Simons level and 
$S_{\rm CS}$ is the Chern-Simons term normalized as
\beq
S_{\rm CS} ~\equiv~ \frac{1}{4\pi} \int {\rm tr} \left[ A \wedge dA + \frac{2i}{3} A \wedge A \wedge A \right].
\eeq

By rescaling the gauge field as $A_\mu \rightarrow A_\mu/k$, 
we can see that in the infinite level limit $k \rightarrow \infty$, 
this model reduces to the free theory
whose equations of motion are 
given by the Schr\"odinger equation.
In addition to the standard Schr\"odinger symmetry, 
the action is invariant under 
the non-relativistic version of superconformal symmetry, 
namely the super-Schr\"odinger symmetry. 
We can show that  
this system has the same symmetry as 
the free supersymmetric Schr\"odinger system 
even for finite $k$. 

\begin{table}[!h]
\begin{center}
\begin{tabular}{ll} ~
$\bullet$ ~$H$ : time translation
& \hs{10}
$\bullet$ ~$P_i$ : translation
\\ ~
$\bullet$ ~$J$ : rotation
& \hs{10}
$\bullet$ ~$D$ : dilatation
\\ ~
$\bullet$ ~$B^i$ : Galilean symmetry
& \hs{10}
$\bullet$ ~$C$ : special Schr\"odinger symmetry
\\ ~
$\bullet$ ~$\mathcal N$ : central charge: phase rotation
& \hs{10}
$\bullet$ ~$\mathcal N_f$ : fermion number symmetry 
\vs{3}
\\ 
\hdashline 
\multicolumn{2}{c}{$\bullet$ ~$Q,~q,~S$ : supersymmetry 
$\phantom{\bigg[}$} 
\end{tabular}
\caption{generators of the super-Schr\"odinger symmetry}
\label{Tab:generators}
\end{center}
\end{table}
\begin{figure}[!h]
\centering
\includegraphics[width=100mm]{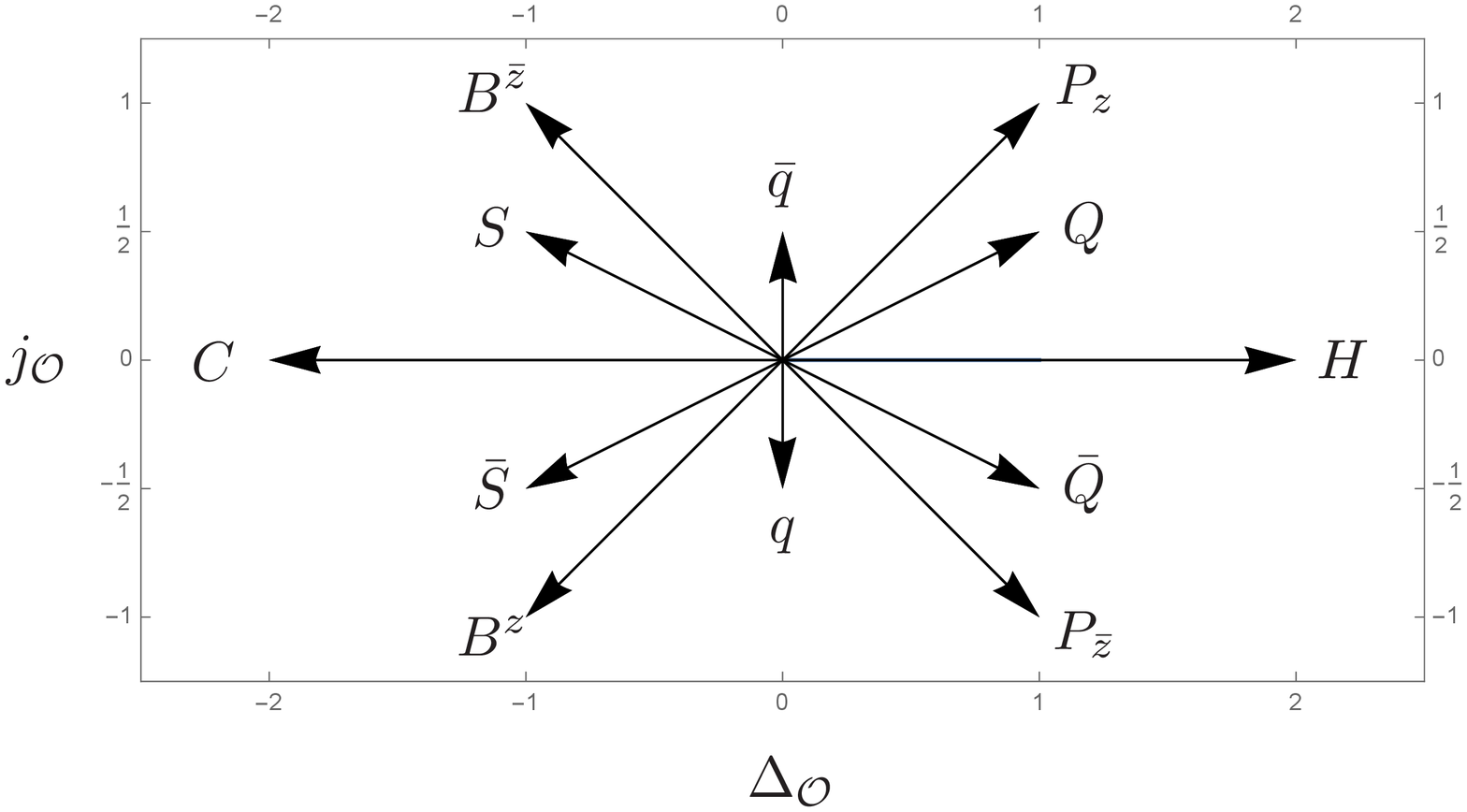}
\caption{$(\Delta_{\mathcal O},\, j_{\mathcal O})$ of the generators}
\label{fig:weight}
\end{figure}

\paragraph{Super-Schr\"odinger algebra \\}
The generators of the super-Schr\"odinger symmetry
is summarized in Table \ref{Tab:generators}.
The non-vanishing part of their commutation relation is given by
\beq
& \displaystyle 
[H,B^z]=-iP_{\bar z}, \hs{5} 
[H,C]=iD, \hs{5}
[P_{\bar z},C]=-iB^z, \hs{5}
[P_{\bar z}, B^{\bar z}]=-2im \mathcal N, 
\phantom{\bigg[} & \\
& \displaystyle 
[J,\mathcal O]= - j_{\mathcal O} \mathcal O, \hs{5}
[D,\mathcal O]=-i\Delta_{\mathcal O} \mathcal O, \hs{5}
[\mathcal N_f, \mathcal O] = -q_f^{\mathcal O} \mathcal O, 
\phantom{\bigg[} &
\eeq
($q_{\mathcal O} = 1$ for $(Q,\,q,\,S)$ and 
$q_{\mathcal O} = 0$ for the bosonic operators
and see Fig.\,\ref{fig:weight} for 
$(\Delta_{\mathcal O}, j_{\mathcal O}$))
\beq
& \displaystyle
\{ Q, \bar Q \} = 2 H, \hs{5} 
\{Q,\bar q \} = P_z, \hs{5} 
\{Q, \bar S \} = D + i J - \frac{3}{2} i \mathcal N_f.,
& \phantom{\bigg]} \\
& \displaystyle
\{ q, \bar q \} = m \mathcal N, \hs{5}
\{q, \bar S \} = - B^z, \hs{5} \{S, \bar S \} = 2 C, 
& \phantom{\bigg]} \\
& \displaystyle
[H,S]=i Q, \hs{5}
[C,Q]=-iS, \hs{5}
[P_{\bar z}, S]=[B^z,Q]=2i q. \hs{5}
& \phantom{\bigg]} 
\eeq

\paragraph{Bosonic part of super-Schr\"odinger symmetry \\}
Let $\xi^\mu$ be the non-relativistic version of 
the conformal Killing vector 
\beq
\xi^t = \varepsilon_H + 2 \varepsilon_D t - \varepsilon_C t^2, \hs{10} 
\xi^z = - 2 (\varepsilon_P + \varepsilon_{\bar B} t) + 
(\varepsilon_D - \varepsilon_C t + i \varepsilon_J ) z, \hs{10}
\xi^{\bar z} = \overline{\xi^z },
\eeq
where $\epsilon_{\mathcal O}$ are transformation parameters.
Then the bosonic part of 
the super-Schr\"odinger transformations 
takes the form 
\begin{align}
\hs{10}
\delta \phi \ &= \Big[ \xi^\mu \D_\mu + \lambda + i \alpha \Big] \phi, 
& \delta \psi \ &= \Big[ \xi^\mu \D_\mu + \lambda + i \alpha + i \varepsilon_f \Big] \psi, \hs{10}
\\
\hs{10}
\delta \phi^\dagger &= \Big[ \xi^\mu \D_\mu + \lambda - i \alpha \Big] \phi^\dagger, 
& \delta \psi^\dagger &= \Big[ \xi^\mu \D_\mu + \lambda - i \alpha - i \varepsilon_f \Big] \psi^\dagger, \hs{10}
\end{align}
\vs{-10}
\beq
\delta A_\mu = \xi^\nu F_{\nu \mu},
\eeq
where the real functions $\lambda$ and $\alpha$ are given by
\beq
\lambda = \varepsilon_D - \varepsilon_C t, \hs{10}
\alpha = \varepsilon_{\mathcal N} 
+ m z \varepsilon_B 
+ m \bar z \varepsilon_{\bar B} 
+ \frac{m}{2}|z|^2 \varepsilon_C.
\eeq

\paragraph{Fermionic part of super-Schr\"odinger symmetry \\}
To see the invariance of the action under the supersymmetry, 
let us first note that
the following transformation does not change the action: 
\begin{align}
\hs{10}
\delta \phi \ &= \phantom{-}
\frac{1}{m} \left( m \zeta_q - 2i \zeta_Q \D_{z} \right) \psi, 
& \delta \psi \ &= - 
\left( m \bar \zeta_q - 2i \bar \zeta_Q \D_{\bar z} \right) \phi, \hs{10} \label{eq:SUSY1} \phantom{\bigg[} 
\\ \hs{10}
\delta \phi^\dagger &= 
-\frac{1}{m} \left( m \bar \zeta_q + 2i \bar \zeta_Q \D_{\bar z} \right) \psi^\dagger, 
& \delta \psi^\dagger &= 
-\left( m \zeta_q + 2i \zeta_Q \D_{z} \right) \phi^\dagger, \hs{10}
\label{eq:SUSY2} \phantom{\bigg[} 
\\ \hs{10}
\delta A_z &= - \frac{2\pi}{km} \bar \zeta_Q \phi \psi^\dagger,
& \delta A_{\bar z} &= \frac{2\pi}{km} \zeta_Q \psi \phi^\dagger,\label{eq:SUSY3} \phantom{\bigg[}
\end{align}
\vs{-7}
\beq
\displaystyle \delta A_t = 
\frac{\pi}{km^2} \left( m \zeta_q + 2i \zeta_Q \D_{z} \right) \psi \, \phi^\dagger + (h.c.),
\label{eq:SUSY4}
\eeq
where $\zeta_q$ and $\zeta_Q$ are 
fermionic SUSY transformation parameters 
corresponding to the supercharges $q$ and $Q$, respectively. 
Actually, there exists one more supersymmetry 
generated by the supercharge $S$
whose transformation law can be obtained 
from \eqref{eq:SUSY1}-\eqref{eq:SUSY4} 
by promoting the transformation parameters 
$\zeta_q$ and $\zeta_Q$
into the following functions 
depending on the coordinates $(t, z, \bar z)$
\beq
\zeta_q = \varepsilon_q - \bar z \varepsilon_S, \hs{10} 
\zeta_Q = \varepsilon_Q + t \varepsilon_S, 
\label{eq:original_zeta}
\eeq
where $(\varepsilon_q,\,\varepsilon_Q,\,\varepsilon_S)$ 
are transformation parameters 
corresponding to the supercharges $(q,Q,S)$
\beq
\delta = \varepsilon_q q + \varepsilon_Q Q + \varepsilon_S S + (h.c.). 
\eeq

\subsection{Harmonic trap and modified super-Schr\"odinger symmetry} 
Now let us put the SUSY Jackiw-Pi system 
in a harmonic trap. 
The harmonic potential term can be introduced 
by adding the Noether charge $C$ 
(corresponding to the special Schr\"odinger transformation)
to the Hamiltonian. 
We can turn on such \textit{generalized} chemical potential terms
by introducing the following external gauge field $A_\mu^{\rm ex}$
as $A_\mu \rightarrow A_\mu + A_\mu^{\rm ex}$: 
\beq
A_\mu^{\rm ex} dx^\mu \, = \, \frac{i}{2} m \tilde \omega (\bar z dz - z d \bar z) + \bigg[ \frac{m}{2} ( \omega^2 - \tilde \omega^2 ) |z|^2 
- \mu_f \, \hat{\mathcal N}_f - \mu_a \, \hat{\mathcal N}_a \bigg] dt,
\eeq
where $\hat{\mathcal N}_f$ is the fermion number operator
\beq
\hat{\mathcal N}_f \, \psi =\psi, \hs{10}  
\hat{\mathcal N}_f \, \phi = 0, 
\eeq 
and $\hat{\mathcal N}_a~(a=1,\cdots,\NF)$ are 
the flavor number operator
\beq
\hat{\mathcal N}_a \, \phi_b = \delta_{ab} \, \phi_b, \hs{10}
\hat{\mathcal N}_a \, \psi_b = \delta_{ab} \, \psi_b.
\eeq
The parameters $(\omega, \tilde \omega, \mu_f , \mu_a)$
correspond to the following \textit{generalized} chemical potentials
\begin{itemize}
 \setlength{\leftskip}{3.0cm}
\item 
$\omega$ ~ : the strength of the harmonic trap 
\item
$\tilde \omega$ ~ : the angular velocity of the rotation 
\item
$\mu_f$ \, : the fermion number chemical potential
\item
$\mu_a$ \, : the flavor symmetry chemical potential 
\end{itemize}

In the presence of the external gauge fields, 
the differential operator $\hat \Delta_0$ in the kinetic terms
is replaced by $\hat \Delta$ obtained by replacing the covariant derivatives with those with the external field 
\beq
\hat \Delta &\equiv& i \tilde{\D}_t + \frac{1}{m} \left( \tilde{\D}_z \tilde{\D}_{\bar z} + \tilde{\D}_{\bar z} \tilde{\D}_z \right) \notag \\
&=& \hat \Delta_0 - \tilde \omega ( z \D_z - \bar z \D_{\bar z} ) - \frac{m \omega^2}{2}|z|^2 + \mu_f \hat{\mathcal N}_f + \sum_{a=1}^{\NF} \mu_a \hat{\mathcal N}_a,
\eeq
where $\tilde{\D}_\mu$ denotes the covariant derivative 
including the external field
\beq
\tilde{\D}_\mu \phi \equiv (\p_\mu + i A_\mu + i A_\mu^{\rm ex}) \phi, ~ etc. 
\eeq

Since the differential operator $\hat \Delta$ does not commute
with some generators of the super-Schr\"odinger transformation, 
it appears that a part of the super-Schr\"odinger symmetry
including the supersymmetry
is broken in the Jackiw-Pi system in the harmonic trap. 
Although the original super-Schr\"odinger transformation
is no longer a symmetry of the action, 
there exists a modified super-Schr\"odinger symmetry
even in the presence of the external field. 

\paragraph{Bosonic part of modified super-Schr\"odinger symmetry \\}
To write down the bosonic part of 
the modified super-Schr\"odinger symmetry
it is convenient to introduce $\eta_I(t)$ 
defined by the following differential equations 
\beq
i \p_t (\eta_{\bar B} \mp i \omega \eta_P) = (\tilde \omega \pm \omega)(\eta_{\bar B} \mp i \omega \eta_P), \hs{10}
i \p_t ( \eta_C + 2 i \omega \eta_D) = 
2 \omega ( \eta_C + 2 i \omega \eta_D), 
\label{eq:mod_Sch1}
\eeq
\beq
\p_t \eta_J = - \tilde \omega \p_t \eta_H = - 2 \tilde \omega \eta_D, 
\hs{10}
\p_t \eta_{\mathcal N} = \p_t \eta_{f} = 0. 
\label{eq:mod_Sch2}
\eeq
By using these functions, 
``the non-relativistic conformal Killing vector $\xi^\mu$
for the modified Schr\"odinger symmetry"
can be written as
\beq
\xi^t = \eta_H, \hs{10} 
\xi^z = - 2 \eta_P + (\eta_D + i \eta_J) z, \hs{10}
\xi^{\bar z} = - 2 \eta_{\bar P} + (\eta_D - i \eta_J) \bar z.  
\eeq
Then the bosonic part of 
the modified super-Schr\"odinger transformations 
takes the form 
\begin{align}
\hs{10}
\delta \phi \ &= \Big[ \xi^\mu \D_\mu + \lambda + i \alpha \Big] \phi, 
& \delta \psi \ &= \Big[ \xi^\mu \D_\mu + \lambda + i \alpha + i \eta_f \Big] \psi, \hs{10}
\\
\hs{10}
\delta \phi^\dagger &= \Big[ \xi^\mu \D_\mu + \lambda - i \alpha \Big] \phi^\dagger, 
& \delta \psi^\dagger &= \Big[ \xi^\mu \D_\mu + \lambda - i \alpha - i \eta_f \Big] \psi^\dagger, \hs{10}
\end{align}
\vs{-7}
\beq
\delta A_\mu = \xi^\nu F_{\nu \mu},
\eeq
with $\lambda = \eta_D$ and 
\beq
\alpha = \eta_{\mathcal N} 
+ m (\eta_B - i \tilde \omega \eta_{\bar P}) z 
+ m (\eta_{\bar B} + i \tilde \omega \eta_P) \bar z 
+ m \left[ \frac{1}{2} \eta_C + \tilde \omega \eta_J - \frac{1}{2}(\omega^2-\tilde \omega^2) \eta_H \right] |z|^2.
\eeq
By appropriately identifying the integration constants 
of the differential equations \eqref{eq:mod_Sch1}-\eqref{eq:mod_Sch2} with the transformation parameters 
$\mathcal \varepsilon_{\mathcal O}$, 
we can confirm that this transformation reduces 
to the standard Schr\"odinger symmetry 
when the chemical potential terms are turned off
($\omega=\tilde \omega=0$). 

It is worth noting that 
the $SU(\NF)$ flavor symmetry is also not broken 
but modified as 
\beq
\delta \phi = i \phi \, T(t), \hs{10} 
\delta \psi = i \psi \, T(t),
\label{eq:mod_flavor}
\eeq
where $T(t)$ denotes a time dependent generator of $SU(\NF)$ 
such that
\beq
i \p_t T(t) = \big[ \mathcal M, T(t) \big], \hs{10} 
T(0) \in \mathfrak{su}(\NF), \hs{10}
\mathcal M \equiv {\rm diag}(\mu_1,\cdots,\mu_{\NF}).
\eeq

\paragraph{Fermionic part of modified super-Schr\"odinger symmetry \\}
The fermionic part of 
the modified super-Schr\"odinger transformation 
takes the same form as the unmodified one  
\eqref{eq:SUSY1}-\eqref{eq:SUSY4}, 
if the covariant derivative is promoted 
as $\D_\mu \rightarrow \tilde{\D}_\mu$ 
and $\zeta_q$ and $\zeta_Q$ are replaced with
the functions satisfying the differential equation
\beq 
\ba{cc} i \p_t - \mu_f & i \bar z (\tilde \omega^2-\omega^2) \\ i \p_{\bar z} & i \p_t - \mu_f  - 2 \tilde \omega \ea \ba{c} \zeta_q \\ \zeta_Q \ea = 0, \hs{10} \p_z \zeta_Q = \p_{\bar z} \zeta_Q = \p_z \zeta_q = 0.
\eeq
The general solution takes the form
\beq
\zeta_q \ = \ e^{- i \mu_f t} \varepsilon_q - e^{- i(\mu_f+2\tilde \omega)t} \, \bar z \, f'(t), \hs{10} 
\zeta_Q \ = \ e^{- i(\mu_f + 2\tilde \omega)t} \, f(t), 
\eeq
where the function $f(t)$ is given by
\beq
f(t) = \frac{1}{2} \left[ \left( \varepsilon_Q + \frac{i \varepsilon_S}{\omega} \right) e^{i(\tilde \omega - \omega)t} + \left( \varepsilon_Q -\frac{i \varepsilon_S}{\omega} \right) e^{i(\tilde \omega + \omega)t} \right],
\eeq
and $f'(t)$ is the time derivative of $f(t)$. 
The integration constants 
($\varepsilon_q, \varepsilon_Q, \varepsilon_S$), 
which can be interpreted as the transformation parameters of 
the modified symmetry, 
are chosen so that $\zeta_q$ and $\zeta_Q$ 
reduce to the original forms \eqref{eq:original_zeta} 
in the limit $\omega,\, \tilde \omega,\, \mu_f \rightarrow 0$.

\section{1/3 BPS equation and Jackiw-Pi vortices}\label{sec:BPS}
\subsection{1/3 BPS condition}
Since the SUSY Jackiw-Pi system
has the modified super-Schr\"odinfer symmetry
even in the harmonic trap,  
it is possible to consider BPS states of the Jackiw-Pi vortices 
\cite{Jackiw:1990tz}
which preserve a part of the modified super-Schr\"odinfer symmetry. 
A BPS condition can be obtained by requiring 
$\delta \psi = 0$ for each choice of the transformation parameters 
$(\varepsilon_q,\varepsilon_Q,\varepsilon_S)$. 
We can obtain a BPS equation 
with no explicit time dependence by setting 
\beq
\varepsilon_q = 0, \hs{10}
\varepsilon_S = - i \omega \epsilon_Q. 
\label{eq:parameter_choice}
\eeq 
By using the differential operators $\nabla_z$ 
and $\nabla_{\bar z}$ defined by
\beq
\nabla_z \equiv \D_z - \frac{1}{2} m \omega \bar z, \hs{10}
\nabla_{\bar z} \equiv \D_{\bar z} + \frac{1}{2} m \omega z, 
\label{eq:nabla_z}
\eeq
we can write the BPS equation corresponding to \eqref{eq:parameter_choice} as
\beq
\nabla_{\bar z} \, \phi = 0. 
\label{eq:BPS}
\eeq
Any solution of this BPS solution satisfies 
the full set of the equations of motion are satisfied 
if the following first order differential equations are 
also satisfied 
\beq
i F_{z \bar z} + \frac{\pi}{k} \phi \phi^\dagger = 0, 
\hs{10}
\nabla_t \, \phi = 0,
\label{eq:Gauss_t}
\eeq
where we have defined 
\beq
\nabla_t \ \equiv \ \D_t + \frac{\pi i}{k m} \phi \phi^\dagger 
+ i (\omega-\tilde \omega) (z \D_z - \bar z \D_{\bar z}) + i \sum_{a=1}^{\NF} (\omega - \mu_a) \hat{\mathcal N}_a.
\eeq
In the following, we consider field configurations 
satisfying the set of equations \eqref{eq:BPS} and \eqref{eq:Gauss_t} with asymptotic behaviors
\beq
\phi \rightarrow 0 , \hs{10} 
F_{\mu \nu} \rightarrow 0.
\eeq

\subsection{General BPS solution}
To write down the general solution of 
the equations \eqref{eq:BPS} and \eqref{eq:Gauss_t} 
it is convenient to introduce 
an arbitrary $N$-by-$\NF$ holomorphic matrix $H_0(t, z)$, 
called the moduli matrix \cite{Eto:2005yh, Eto:2006pg}. 
By using $H_0(t, z)$, 
we can formally solve the equations \eqref{eq:BPS} and \eqref{eq:Gauss_t} as
\beq
\phi \, &=& \, 
S^{-1} H_0(t, z) \, 
{\rm diag} (e^{i \mu_1 t}, e^{i \mu_2 t}, \cdots, e^{i \mu_{\NF}}), \phantom{\bigg(} \label{eq:sol_1} \\
A_{\bar z} &=& 
\hs{-2} -i S^{-1} \p_{\bar z} S 
+ \frac{i}{2} m \omega z , 
\phantom{\bigg(} \label{eq:sol_2} \\
A_t &=& 
\hs{-2} - \frac{\pi}{k m} \phi \phi^\dagger 
+ i (\omega - \tilde \omega) (z A_z - \bar z A_{\bar z})
- \omega
, \phantom{\bigg(} 
\label{eq:sol_3}
\eeq 
where the $N$-by-$N$ matrix $S(t, z, \bar z)$ is 
an element of the complexified gauge group 
$U(N)^\C \cong GL(N,\C)$ satisfying
\beq
\p_{\bar z} ( \p_z \Omega \Omega^{-1}) ~ = ~ 
m \omega - \frac{\pi}{k} H_0 H_0^\dagger \Omega^{-1}, \hs{10} 
\Omega \equiv S S^\dagger. 
\label{eq:master}
\eeq
This equation ensures that the Gauss law 
$i F_{z \bar z} + \frac{\pi}{k} \phi \phi^\dagger$ is satisfied. 
The BPS equation $\nabla_{\bar z} \phi = 0$, 
which can be rewritten as 
\beq
\p_{\bar z} H_0(t,z) = 0,
\eeq
is automatically satisfied for an arbitrary choice of 
the holomorphic matrix $H_0(t,z)$. 
The remaining equation $\nabla_t \phi = 0$ determines 
the time dependence of the solution as 
\beq
\Big[ \p_t + i (\omega-\tilde \omega) (z \p_z - \bar z \p_{\bar z}) \Big] S = \Big[ \p_t + i (\omega-\tilde \omega) (z \p_z - \bar z \p_{\bar z}) \Big] H_0 = 0.
\eeq
These equations imply that $S$ and $H_0$ have 
no explicit $t$-dependence
if they are written in terms of the coordinates 
$z_\ast$ and $\bar z_\ast$ defined as
\beq
z_\ast \equiv e^{i(\tilde \omega - \omega)t} \, z, \hs{10}
\bar z_\ast \equiv e^{-i(\tilde \omega - \omega)t} \, \bar z. 
\eeq
This implies that the whole system is rotating in the $z$-plane 
with angular velocity $\tilde \omega - \omega$. 
By solving the equation \eqref{eq:master} for $\Omega$, 
physical quantities such as energy density profiles
can be explicitly obtained for an arbitrarily chosen $H_0(z_\ast)$. 
Note that for $N=\NF=1$, Eq.\,\eqref{eq:master} can be rewritten 
into the vortex equation classified as follows \cite{Manton:2016waw}
\begin{align}
m\omega &= 0, ~k < 0 \hs{-35} & \mbox{Jackiw-Pi \cite{Jackiw:1990tz}} \notag \\
m\omega &> 0, ~k < 0 \hs{-35} & \mbox{Ambj\o rn-Olesen \cite{Ambjorn:1988fx, Ambjorn:1988tm}} \notag \\
m\omega &> 0, ~k > 0  \hs{-35} & \mbox{Taubes \cite{Taubes:1979tm}} \notag \\
m\omega &> 0, ~k = \infty \hs{-35} & \mbox{Bradlow \cite{Bradlow:1990ir}} \notag \\
m\omega &< 0, ~k < 0 \hs{-35} & \mbox{Popov \cite{Popov:2012av}} \notag
\end{align}
Although Eq.\,\eqref{eq:master} has 
the identical form with the vortex equation, 
the boundary conditions are different and consequently 
the vortices in our setup have some distinctive physical properties.

The stability of the solution is guaranteed by 
the conserved charges associated with 
the spatial rotation and the internal phase rotation. 
We can show that for given values of the Noether charges, 
the energy of the system is bounded from below as
 \beq
E \ \geq \ (\omega - \tilde \omega) J + \sum_{a=1}^{\NF} (\omega -\mu_a) \mathcal N_a,
\eeq
where $J$ and $\mathcal N_a$ are the angular momentum and 
the flavor symmetry Noether charges 
\beq
J = \int d^2 x \, \phi_a^\dagger (z \D_z - \bar z \D_{\bar z}) \phi_a, 
\hs{10}
\mathcal N_a = \int d^2 x \, \phi_a^\dagger \phi_a~~(\mbox{no sum over $a$}). 
\eeq
The BPS solution saturates this lower bound for the energy.
This is an example of $Q$ solitons,
that is, solitons which are stabilized by Noether charges.  

It is worth noting that 
we can obtain more general solutions of the equations of motion
(breathing solutions, etc) 
by applying the modified Schr\"odinger transformation
to the BPS configurations discussed in this section. 
Such solutions preserve different combination of the supercharges 
and satisfy a certain time-dependent BPS equations. 

\subsection{Static BPS solution}
\begin{figure}[!h]
\begin{center}
\begin{minipage}[h]{0.45\hsize}
\begin{center}
\includegraphics[width=70mm, bb = 43 35 755 529]{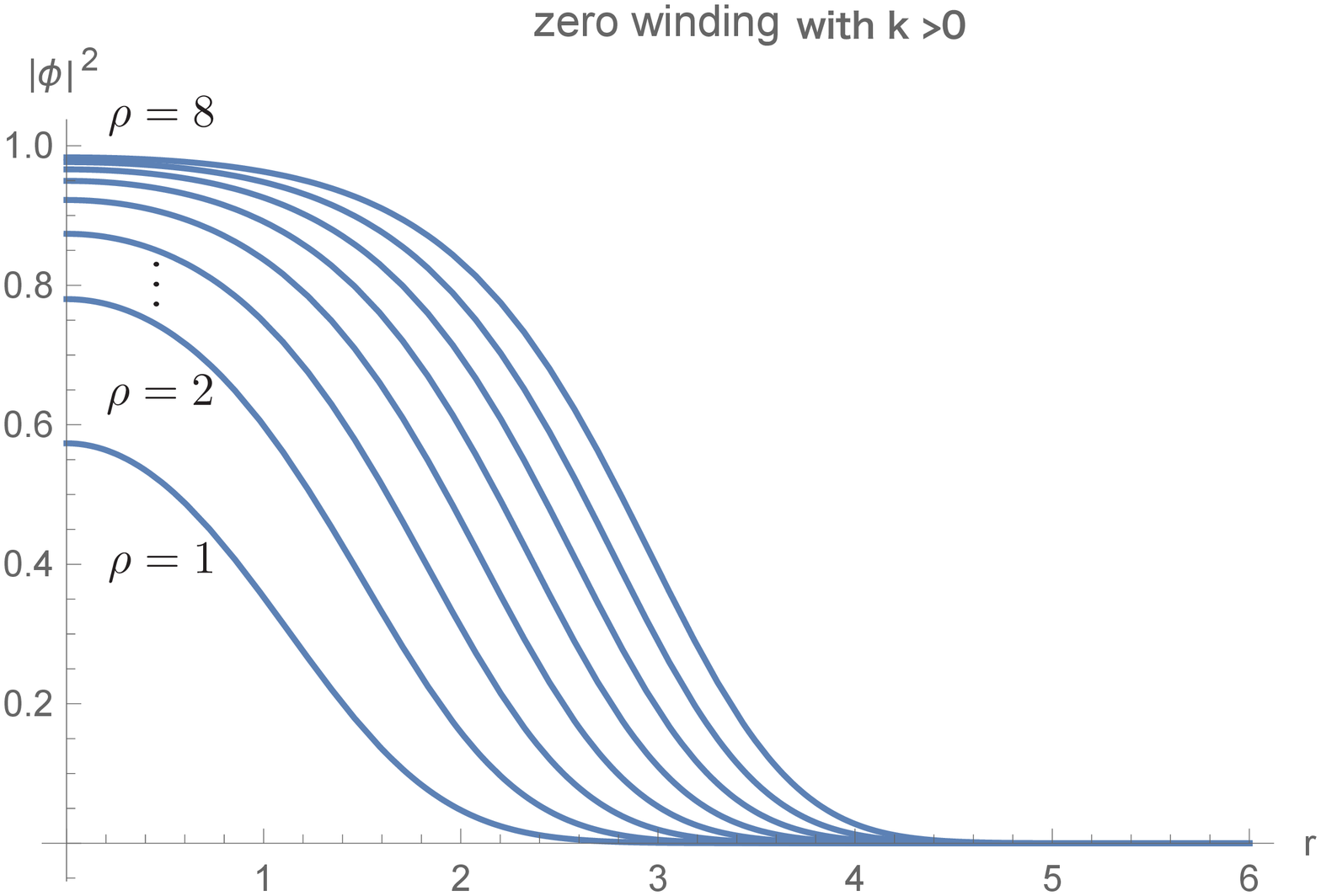} \\
$\phantom{\bigg[}$ (a) $k>0$, $l=0$
\end{center}
\end{minipage}
\begin{minipage}[h]{0.45\hsize}
\begin{center}
\includegraphics[width=70mm, bb = 43 35 755 529]{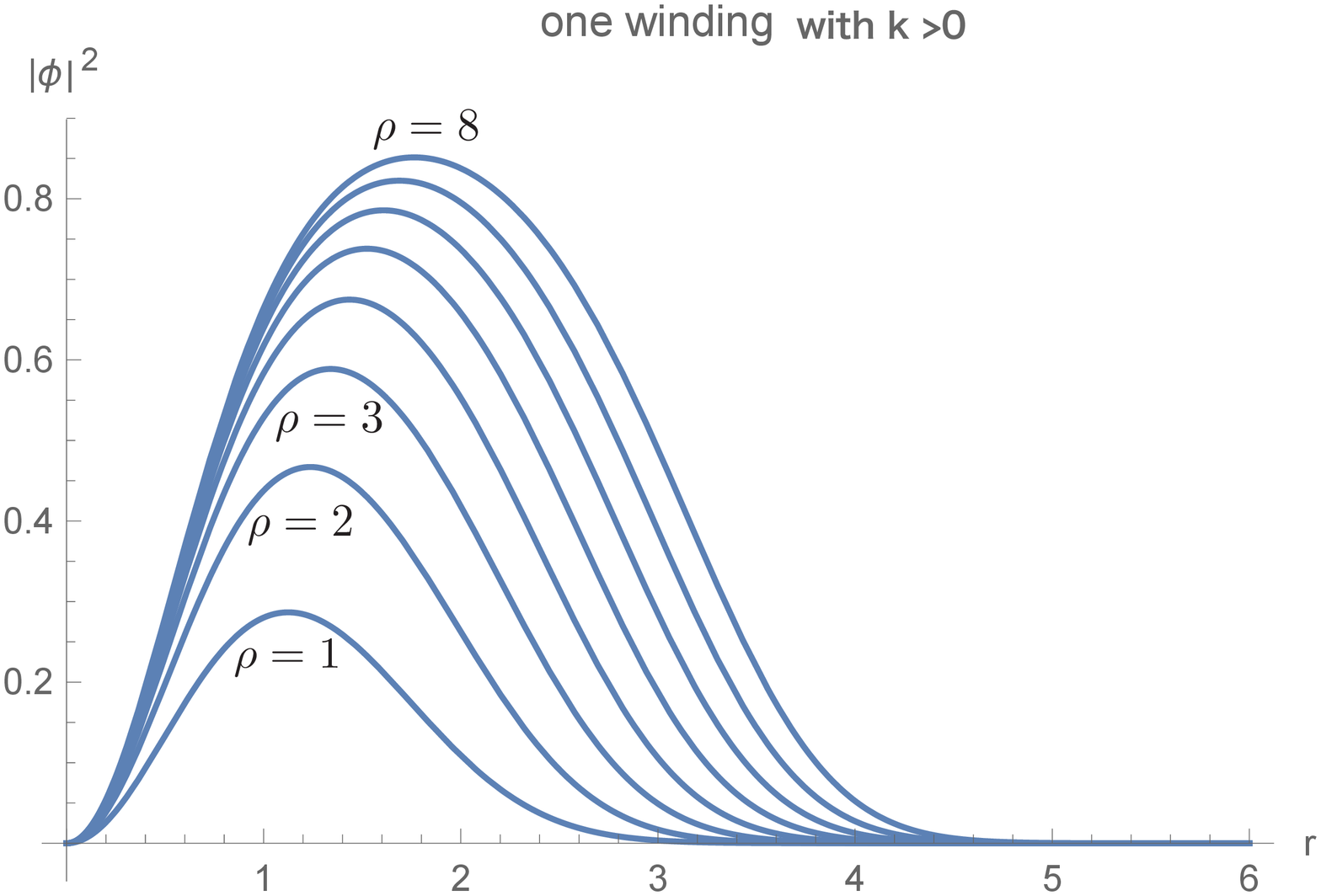} \\ 
$\phantom{\bigg[}$ (b) $k>0$, $l=1$
\end{center}
\end{minipage}
\hs{7}
\caption{Charge density profiles for $k=\pi$, $N=\NF=1$, $m\omega=1$.}
\label{fig:posi}
\end{center}
\end{figure} 
\begin{figure}[!h]
\begin{center}
\begin{minipage}[h]{0.45\hsize}
\begin{center}
\includegraphics[width=70mm, bb = 43 35 755 529]{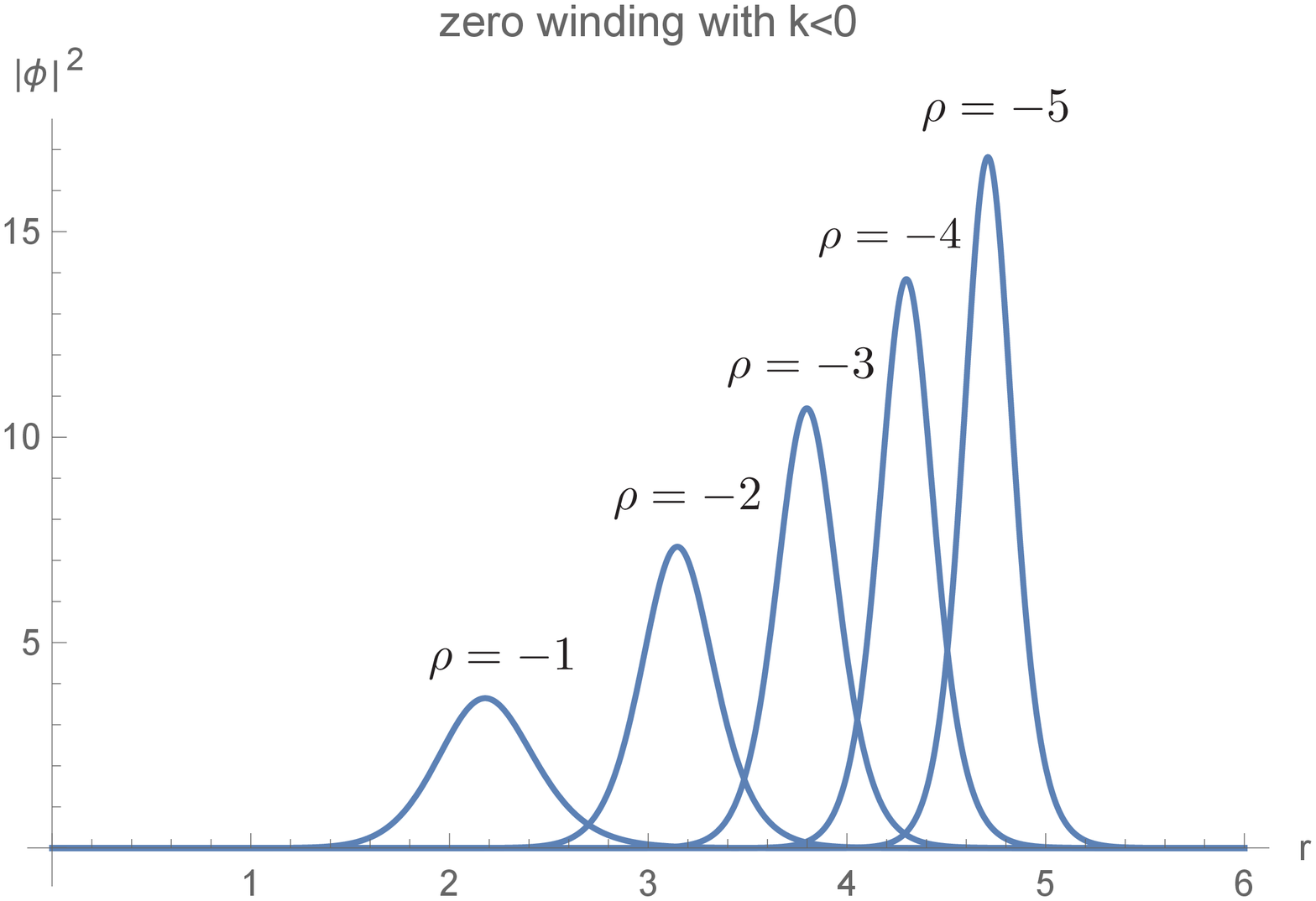} \\
$\phantom{\bigg[}$ (a) $k<0$, $l=0$
\end{center}
\end{minipage}
\begin{minipage}[h]{0.45\hsize}
\begin{center}
\includegraphics[width=70mm, bb = 43 35 755 529]{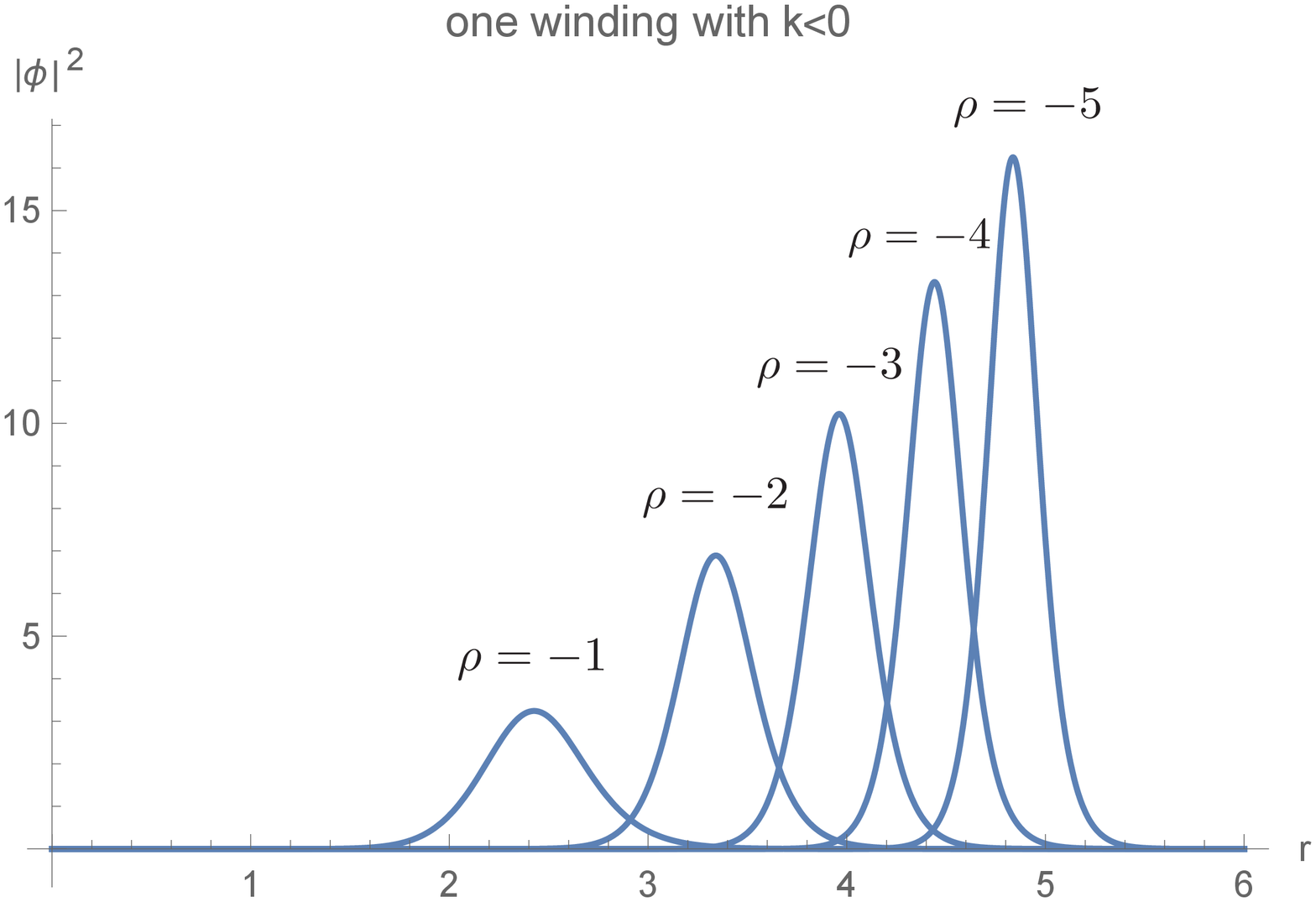} \\ 
$\phantom{\bigg[}$ (b) $k<0$, $l=1$
\end{center}
\end{minipage}
\hs{7}
\caption{Charge density profiles for $k=-\pi$, $N=\NF=1$, $m\omega=1$.}
\label{fig:nega}
\end{center}
\end{figure} 
Although a generic BPS configuration 
is a stationary solution which depends on time $t$, 
the solution \eqref{eq:sol_1}-\eqref{eq:sol_3} becomes static 
if $H_0$ is chosen so that 
the resulting scalar field $\phi$ is invariant under 
the rotation and the flavor transformations, 
that is
\beq
\hat J \, \phi \ = \ \hat{\mathcal N}_a \, \phi \ = \ 
0 + \{ \mbox{infinitesimal gauge transformation} \}. 
\eeq
By appropriately fixing the gauge, 
the static solution with e.g. 
$\mu_a \, \hat{\mathcal N}_a \, \phi 
= \boldsymbol \mu \phi \equiv {\rm diag}(\mu_1,\cdots \mu_N) \phi$, 
can be written as
\beq
\phi \ &=& \, \Big( e^{-\frac{1}{2} \boldsymbol \sigma} z^{\boldsymbol L} \, \Big| \, \mathbf 0_{N \times (\NF-N)} \Big), \phantom{\bigg(} \label{eq:static_sol_1} \\
A_{\bar z} &=& 
\hs{-2} - \frac{i}{2} \p_{\bar z} \, 
\left( \boldsymbol \sigma - m \omega |z|^2  \right), 
\phantom{\bigg(} \label{eq:static_sol_2} \\
A_t &=& 
\hs{-2} - \frac{\pi}{k m} \phi \phi^\dagger 
+ i (\omega - \tilde \omega) 
(z A_z - \bar z A_{\bar z} + i \boldsymbol L)
- (\omega - \boldsymbol \mu), 
\phantom{\bigg(} 
\label{eq:static_sol_3}
\eeq
where $\boldsymbol L = {\rm diag} (l_1, l_2,\cdots, l_N)$ 
is an $N$-by-$N$ diagonal matrix with $l_i \in \Z_{\geq 0}$. 
The matrix $\boldsymbol \sigma = {\rm diag}(\sigma_1, \sigma_2, \cdots, \sigma_N)$ denotes a set of real profile functions satisfying
\beq
\p_z \p_{\bar z} \sigma_i = m \omega - \frac{\pi}{k} |z|^{2l_i} e^{-\sigma_i}, 
\hs{10}
(z \p_z - \bar z \p_{\bar z}) \sigma_i = 0, 
\eeq
with asymptotic behavior $\sigma_i \rightarrow m \omega |z|^2$. 
We can show that the subleading part of $\sigma_i$ takes the form
\beq
\sigma_i \rightarrow m \omega |z|^2 - \rho_i \log |z|^2,
\eeq
From this asymptotic behavior, it follows that
the real parameters $\rho_i$ correspond to 
the magnetic fluxes and the flavor charges
\beq
\frac{1}{2\pi} \int d^2 x \, i F_{z \bar z} = 
- \frac{1}{2} {\rm diag} ( \rho_1, \cdots, \rho_N ), \hs{10} 
\mathcal N_a = \frac{k}{2\pi} \rho_a.
\eeq
See Figs.\,\ref{fig:posi} and \ref{fig:nega} 
for the profiles of the charge density.

As was done in \cite{Ezawa:1991xf}, 
time-dependent solutions can be obtained 
from these static static solutions 
by applying the modified Schr\"odinger symmetry. 
Although such solutions do not satisfy 
the BPS equation \eqref{eq:BPS}, 
they preserve a certain time-dependent 
linear combination of the supercharges. 

\section{Spectrum of fluctuation modes in BPS background}
\label{sec:fluc}
In this section we consider fluctuations of the fields
$(\delta A_\mu, \delta \phi, \delta \psi)$ 
around a BPS background $(A_\mu, \phi)$
and show that in addition to 
the so-called massive Nambu-Goldstone modes 
in the bosonic fluctuations, 
there exist fermionic massive Nambu-Goldstone modes
associated with the broken supercharges of 
the modified super-Schr\"odinger symmetry. 

\subsection{Linearized equations for fluctuations}
When we discuss fluctuations of the bosonic fields, 
it is convenient to remove the gauge zero modes
by imposing the gauge fixing condition on the fluctuations as
\beq
\delta A_t = - \frac{\pi}{k m} (\delta \phi \phi^\dagger + \phi \delta \phi^\dagger) - i (\omega-\tilde \omega)(z \delta A_z - \bar z \delta A_{\bar z}). 
\label{eq:linear_gauge}
\eeq
This gauge fixing condition does not completely remove 
unphysical gauge zero modes since 
there still exist remaining gauge degrees of freedom
generated by $\Lambda \in \mathfrak u(N)$ such that
\beq
\D_t \Lambda + i (\omega-\tilde \omega) (z \D_z \Lambda - \bar z \D_{\bar z} \Lambda) + \frac{\pi i}{km} [\phi \phi^\dagger, \Lambda] = 0.
\label{eq:residual}
\eeq
This residual gauge degrees of freedom can be fixed 
by imposing the additional gauge fixing condition as
\beq
i ( \D_z \delta A_{\bar z} + \D_{\bar z} \delta A_z ) + \frac{\pi}{k} (\delta \phi \phi^\dagger - \phi \delta \phi^\dagger) = 0.
\label{eq:additional_cond}
\eeq
The Gauss law equation 
$i F_{z \bar z} + \frac{\pi}{k} \phi \phi^\dagger = 0$ 
reduces to the linearized Gauss law 
for the fluctuation fields
\beq
i ( \D_z \delta A_{\bar z} - \D_{\bar z} \delta A_z ) + \frac{\pi}{k} (\delta \phi \phi^\dagger + \phi \delta \phi^\dagger) = 0. 
\label{eq:linear_gauss}
\eeq
If the gauge fixing condition \eqref{eq:linear_gauge}
and the linearized Gauss law \eqref{eq:linear_gauss} are satisfied, 
the linearized equations for the fluctuation fields can be written as
\beq
\left( i \nabla_t+ \frac{2}{m} \boldsymbol{\nabla} \tilde{\boldsymbol{\nabla}} \right)
\ba{c}
\delta A_{\bar z} \\
\delta \phi
\ea
=0 ,
\hs{10}
\left( i \nabla_t +\frac{2}{m} \tilde{\boldsymbol{\nabla}} \boldsymbol{\nabla} \right) \delta \psi = 0,
\label{eq:linearaized_eq}
\eeq
where the differential operators are given by\footnote{
The generator $\hat J$ denotes
the angular momentum operator including the spin part
\beq
\hat J \equiv z \D_z - \bar z \D_{\bar z} + \hat S. \notag
\eeq
}
\beq
\nabla_t \ \equiv \ \D_t + \frac{\pi i}{k m} \phi \phi^\dagger 
+ i (\omega-\tilde \omega) \left( \hat J - \frac{3}{2} \hat{\mathcal N}_f \right) + i \sum_{a=1}^{\NF} (\omega - \mu_a) \hat{\mathcal N}_a - i (\mu_f+\omega+\tilde \omega) \hat{\mathcal N}_f,
\label{eq:nabla_t}
\eeq
and
\beq
\boldsymbol{\nabla} = 
\ba{c}
\frac{\pi i}{k} \hat \phi^\dagger \\
\nabla_z 
\ea,
\hs{10}
\tilde{\boldsymbol{\nabla}} = 
\ba{cc}
i \hat \phi & \nabla_{\bar z} 
\ea. 
\eeq
The operators $\hat \phi$ and $\hat \phi^\dagger$ denote 
the right multiplications of $\phi$ and $\phi^\dagger$, 
e.g. $\hat \phi \cdot \delta A_{\bar z} = \delta A_{\bar z} \phi_a$, 
and the differential operators $\nabla_z$ and $\nabla_{\bar z}$ 
are defined by
\beq
\nabla_z \equiv \D_z - \frac{1}{2} m \omega \bar z, \hs{10}
\nabla_{\bar z} \equiv \D_{\bar z} + \frac{1}{2} m \omega z. 
\eeq

\subsection{Eigenmode expansion and supermultiplets}
Here we consider fluctuations around the static BPS background \eqref{eq:static_sol_1}-\eqref{eq:static_sol_3}. 
Let us consider the eigenmode expansion of the fluctuations 
\beq
\ba{c}
\delta A_{\bar z} \\
\delta \phi
\ea 
= 
\sum_n 
\varphi_n(t)
\ba{c}
u_{g,\,n} \\
u_{s,\,n}
\ea, 
\hs{10}
\delta \psi = \sum_n \chi_n(t) \, u_{f,\,n}, 
\eeq
where $(u_{g,\,n}, u_{s,\,n})$ and $u_{f,\,n}$ are 
bosonic and fermionic mode functions
satisfying the eigenmode equations 
\beq
\left( i \nabla(\epsilon_{b,\,n}) + \frac{2}{m} \boldsymbol{\nabla} \tilde{\boldsymbol{\nabla}} \right)
\ba{c}
u_{g,\,n} \\
u_{s,\,n}
\ea
=0 ,
\hs{10}
\left( i \nabla(\epsilon_{f,\,n}) +\frac{2}{m} \tilde{\boldsymbol{\nabla}} \boldsymbol{\nabla} \right) u_{f,\,n} = 0,
\eeq
where $\nabla(\epsilon)$ is the operators 
which can be obtained from $\nabla_t$
in \eqref{eq:nabla_t} by replacing the time derivative $i \p_t$
with an eigenvalue $\epsilon$. 
Then the linearized equations reduce to 
the following equations for 
the bosonic and fermionic degrees of freedom $\varphi_n(t)$ and $\chi_n(t)$ 
\beq
i \p_t \varphi_n(t) = \epsilon_{b,\,n} \, \varphi_n(t), \hs{10}
i \p_t \chi_n(t) = \epsilon_{f,\,n} \, \chi_n(t). 
\eeq

Note that since $\nabla(\epsilon)$ commutes with $\boldsymbol{\nabla} \tilde{\boldsymbol{\nabla}}$ and 
$\boldsymbol{\nabla} \tilde{\boldsymbol{\nabla}}$
\beq
\left[ \nabla(\epsilon_{b,\,n}) ,\boldsymbol{\nabla} \tilde{\boldsymbol{\nabla}} \right] 
\ba{c}
u_{g,\,n} \\
u_{s,\,n}
\ea 
= 0, \hs{10}
\Big[ \nabla(\epsilon_{f,\,n}) , \tilde{\boldsymbol{\nabla}} \boldsymbol{\nabla} \Big] u_{f,\,n} = 0, 
\eeq
the solution to the bosonic (fermionic) linearized equation can be decomposed into simultaneous eigenmodes of $\nabla(\epsilon)$ and $\boldsymbol{\nabla} \tilde{\boldsymbol{\nabla}}$ 
($\tilde{\boldsymbol{\nabla}} \boldsymbol{\nabla}$).

Since the BPS background configuration preserves 
a linear combination of three complex supercharges $(q, Q, S)$, 
eigenmodes of the bosonic and fermionic fluctuations are paired 
so that they form supermultiplets of 
the unbroken supersymmetry. 
Let $(u_g, u_s)$ be a bosonic eigenmode 
with eigenvalue $\epsilon_b$. 
The partner fermionic eigenmode $u_f$ can be obtained as
\beq
u_f = \tilde{\boldsymbol \nabla} 
\ba{c}
u_g \\
u_s 
\ea , \hs{10}
\epsilon_f = \epsilon_b - \mu_f - \omega - \tilde \omega. 
\label{eq:b_to_f}
\eeq
On the other hand, any fermionic eigenmode $u_f$ 
with eigenvalue $\epsilon_f$ 
can be mapped to its partner bosonic eigenmode as
\beq
\ba{c}
u_g \\
u_s
\ea 
=
\boldsymbol{\nabla} u_f, 
\hs{10}
\epsilon_b = \epsilon_f + \mu_f + \omega + \tilde \omega.  
\label{eq:f_to_b}
\eeq

Since $(u_g, u_s)$ and $u_f$ are eigenmodes of 
$\boldsymbol{\nabla} \tilde{\boldsymbol{\nabla}}$
and $\tilde{\boldsymbol{\nabla}} \boldsymbol{\nabla}$ respectively,  
the sequential mappings 
($boson \rightarrow fermion \rightarrow boson$) 
and ($fermion \rightarrow boson \rightarrow fermion$) 
do not give new eigenmodes 
\beq
\ba{c}
u_g \\
u_s
\ea 
~\rightarrow~ 
\boldsymbol{\nabla} \tilde{\boldsymbol{\nabla}} 
\ba{c}
u_g \\
u_s
\ea 
~\propto~
\ba{c}
u_g \\
u_s
\ea, 
\hs{10}
u_f ~\rightarrow~ \tilde{\boldsymbol{\nabla}} \boldsymbol{\nabla} 
u_f ~\propto~ u_f.  
\eeq 
Therefore, a generic supermultiplet consists of 
a pair of bosonic and fermionic eigenmodes.

It is worth noting that 
unlike the case of ordinary supersymmetry, 
bosonic and fermionic eigenmodes in a supermultiplet 
have different eigenfrequencies 
$\epsilon_b - \epsilon_f = \mu_f + \omega + \tilde \omega$.
This is due to the unbroken supersymmetry 
is a part of the modified supersymmetry 
which explicitly depends on time $t$. 
One can check that the unbroken supersymmetry 
becomes independent of $t$ when 
$\epsilon_b - \epsilon_f = \mu_f + \omega + \tilde \omega = 0$. 

\paragraph{Short supermultiplets \\}
Although a generic supermultiplet 
made up of a pair of bosonic and fermionic eigenmodes,
there also exist short supermultiplets, 
each of which consists of only a single bosonic eigenmode. 
Such a short multiplet can be found by 
solving the linearized BPS equation
$\nabla_{\bar z} \delta \phi + i \delta A_{\bar z} \phi = 0$, i.e.
\beq
\tilde {\boldsymbol \nabla} 
\ba{c}
u_g \\
u_s
\ea 
= 0.
\label{eq:linear_BPS}
\eeq
For a bosonic eigenmode satisfying this equation, 
the $boson \rightarrow fermion$ 
mapping \eqref{eq:b_to_f} vanishes. 
Furthermore, there is no fermionic eigenmode such that
$\boldsymbol \nabla u_f$ is 
a solution of the linearized BPS equation
since the operator 
$\tilde{\boldsymbol \nabla} \boldsymbol \nabla 
= \nabla_z \nabla_{\bar z} - m \omega$
has no zero mode.\footnote{
$\tilde{\boldsymbol \nabla} \boldsymbol \nabla 
= \nabla_z \nabla_{\bar z} - m \omega$
is a negative definite operator, since for any function $f$, 
\beq
\int d^2 x \, \bar f \left( \nabla_z \nabla_{\bar z} - m \omega \right) f 
\ = \ - \int d^2 x \, \left[ \ \left| \left( \p_{\bar z} + \frac{m \omega}{2} z \right) f \right|^2 + m \omega |f|^2 \, \right] 
\ < \ 0. \notag
\eeq
}
Therefore, this type of the short multiplet 
consists of only a single bosonic mode. 

\paragraph{Linearized Gauss law and gauge fixing condition \\}
Since the bosonic component of a long supermultiplet 
is an element of ${\rm im} \nabla$, 
the linearized Gauss law equation \eqref{eq:linear_gauss}
and the gauge fixing condition \eqref{eq:additional_cond}
are automatically satisfied 
\beq
i \D_z u_g + \frac{\pi}{k} u_s \phi^\dagger \ = \ 
i \D_z \left( \frac{\pi i}{k} u_f \phi^\dagger \right) + \frac{\pi}{k} \nabla_z u_f \phi^\dagger 
\ = \ 0.
\eeq

On the other hand, 
for any solution of the linearized BPS equation 
\eqref{eq:linear_BPS} (element of ${\rm Ker} \, \tilde \nabla$), 
we can always find a short multiplet satisfying
the constraints \eqref{eq:linear_gauss} and \eqref{eq:additional_cond}
by using the symmetry of the eigenmode equation
\beq
\ba{c}
u_g \\
u_s 
\ea
\rightarrow
\ba{c}
u_g - \D_{\bar z} \Lambda \\
u_s + i \Lambda \phi
\ea, 
\label{eq:sym_short}
\eeq
where $\Lambda \in \mathfrak{gl} (N)$ is an $N$-by-$N$ matrix 
satisfying $\nabla(\epsilon_b) \, \Lambda = 0$ and
\beq
i \left[ \D_z \D_{\bar z} \Lambda - \frac{\pi}{k} \Lambda \phi \phi^\dagger \right] \ = \ i \D_z u_g + \frac{\pi}{k} u_s \phi^\dagger.
\eeq
In this way, we can find physical short multiplets 
satisfying the linearized Gauss law equation and 
the gauge fixing condition. 

\subsection{Bosonic and Fermionic massive Nambu-Goldstone modes}
Since the static BPS configuration 
\eqref{eq:static_sol_1}-\eqref{eq:static_sol_3} breaks 
a part of the  super-Schr\"odinger symmetry, 
there exist Nambu-Goldstone (NG) modes 
in the fluctuations of the fields.

\paragraph{Bosonic massive Nambu-Goldstone modes \\}
In the presence of the external fields, 
the super-Schr\"odinger symmetry is modified 
in such a way that the generators explicitly depend on time $t$. 
Consequently, the corresponding NG modes become massive. 
The bosonic NG modes satisfying the constraints \eqref{eq:linear_gauss} and \eqref{eq:additional_cond}
takes the form 
\begin{align}
P_z - i \omega B^{\bar z} 
& ~\rightarrow~
\ba{c}
u_g \\
u_s
\ea 
=
\ba{c}
-i F_{z \bar z} - m \omega \\
- i \nabla_z  \phi
\ea,
&\epsilon_b&= \tilde \omega - \omega, \hs{10} 
\\
P_z + i \omega B^{\bar z} 
&~\rightarrow~
\ba{c}
u_g \\
u_s
\ea 
=
\ba{c}
-i F_{z \bar z} \\
- i \nabla_z \phi
\ea,
&\epsilon_b& = \tilde \omega + \omega. \hs{10}
\\
D + 2 i \omega C 
&~\rightarrow ~  
\ba{c}
u_g \\
u_s
\ea 
=
\ba{c}
z F_{z \bar z} \\
\left( z \D_z + \bar z \D_{\bar z} + 1 \right) \phi
\ea, 
&\epsilon_b&= 2\omega, \hs{10}
\end{align}
where the first NG mode generated by $P_z - i \omega B^{\bar z}$ 
is in a short multiplet and 
we have used the symmetry \eqref{eq:sym_short} 
so that it satisfies the constraint 
\eqref{eq:linear_gauss} and \eqref{eq:additional_cond}. 
These three complex modes 
(and their complex conjugate)
correspond to the broken modified symmetry 
generated by six real operators 
(translation, Galilean, dilatation and special conformal symmetry). 
There also exist massive NG modes corresponding 
to the broken modified flavor symmetry \eqref{eq:mod_flavor}.

\paragraph{Fermionic massive Nambu-Goldstone modes \\}
Since the BPS configuration breaks 
a part of the supersymmetry, 
there also exist fermionic NG modes. 
As in the bosonic case, 
the modified supersymmetry transformations 
explicitly depend on time
and hence the corresponding fermionic NG modes are massive. 
There are two such fermionic massive NG modes 
corresponding to the broken fermionic generators 
$q$ and $Q + i \omega S$
\begin{align}
\hs{10}
q ~~~~
&~\rightarrow~~
u_f = \phi, 
&\epsilon_f& = - \mu_f, \hs{10} 
\label{eq:massive_q} \\
\hs{10}
Q + i \omega S 
&~\rightarrow~~ 
u_f = z \phi, 
&\epsilon_f& = - \mu_f - \tilde \omega + \omega. \hs{10}
\label{eq:massive_QS}
\end{align}
We can check that the NG modes generated by 
$( P_z + i \omega B^{\bar z} , q )$ and 
$( D + 2 i \omega C , Q + i \omega S )$ 
are the pairs of supermultiplets 
related by the boson-fermion mapping discussed above. 

\subsection{Infinite towers of eigenmodes in static BPS background}
As we have seen in the previous section, 
the bosonic and fermionic massive NG modes 
have eigenfrequencies given by 
the chemical potentials 
with the integer coefficients determined 
by the charges of the corresponding generators. 
Here we show that there are 
infinite towers of eigenmodes with such eigenvalues. 

Let us first consider the case of short multiplets. 
Since the BPS equation $\nabla_{\bar z} \phi$ 
is satisfied by \eqref{eq:sol_1}-\eqref{eq:sol_3} 
for an arbitrary matrix $H_0(t,z)$, 
the linearized BPS equation can be solved 
by using the linearized version of \eqref{eq:sol_1}-\eqref{eq:sol_3}, 
which takes the form
\beq
\ba{c}
u_g \\
u_s
\ea 
=
\ba{c}
0 \\
e^{-\frac{1}{2} \boldsymbol \sigma} \delta H_0(z)
\ea 
+
\ba{c}
- \D_{\bar z} \Lambda \\
i \Lambda \phi
\ea,
\eeq
where $\Lambda$ is the $N$-by-$N$ matrix 
determined by the constraint\footnote{
The matrix $\Lambda$ can also be written as 
$\Lambda = i e^{-\frac{1}{2} \boldsymbol \sigma} \delta S$ 
by using the solution $\delta S$ of 
the linearized version of \eqref{eq:master}
which determines the matrix $S$.
}
\beq
i \left[ \D_z \D_{\bar z} \Lambda - \frac{\pi}{k} \Lambda \phi \phi^\dagger \right] = \frac{\pi}{k} e^{-\frac{1}{2} \boldsymbol \sigma} \delta H_0(z) \phi^\dagger. 
\eeq
This solution of the linearized BPS equation 
satisfies the eigenmode equation with eigenvalue $\epsilon$ 
if the matrix $\delta H_0(z)$ is chosen so that 
$\Lambda(\epsilon) \, \delta H_0 = 0$.  
This condition is satisfied when $\delta H_0$ 
has one non-zero component given by a monomial of $z$. 
For example, if $\delta H_0$ has 
$z^l~(l \in \Z_{\geq 0})$ in its $(j,J)$-component 
\beq
(\delta H_0)_{iI} = z^l \delta_{ij} \delta_{IJ}, 
\label{eq:LLL_b}
\eeq
the eigenfrequency of the corresponding short multiplet
is given by  
\beq
\epsilon_b 
= ( l - l_j ) (\omega-\tilde \omega) - \mu_J + \mu_j. 
\eeq
The NG mode generated by $P_z - i \omega B^{\bar z}$ 
corresponds to the linear combination of the modes 
with $(l,J) = (l_j - 1, j)$.
The NG modes corresponding 
to the broken (modified) flavor symmetry 
are also contained in these towers of eigenmodes. 

In addition to the short multiplets, 
we can also find exact spectra of 
a class of ordinary long supermultiplets. 
Such supermultiplets can be obtained from 
the fermionic eigenmode corresponding to the lowest Landau level.
For example, $u_f$ is an eigenmode with frequency
\beq
\epsilon_f = ( l - l_j )(\omega - \tilde \omega) - \mu_J + \mu_j - \mu_f, 
\eeq
if $u_f$ has a non-zero monomial 
in the $(j,J)$-component 
\beq
(u_f)_{iI} = e^{-\frac{1}{2} \sigma_j} z^l \delta_{ij} \delta_{IJ}. 
\label{eq:LLL_f}
\eeq
The fermionic massive NG modes 
\eqref{eq:massive_q} and \eqref{eq:massive_QS}
corresponds to the linear combinations 
of the eigenmodes with $(l,J)=(l_j,j)$ and $(l,J) = (l_j+1,j)$, 
respectively.
The corresponding bosonic mode can be obtained 
by applying the map \eqref{eq:f_to_b}
\beq
(u_g)_{iI} = \frac{\pi i}{k} e^{-\frac{1}{2}(\sigma_j + \sigma_J)}
z^l \bar z^{l_J} \delta_{ij} \delta_{IJ}, \hs{10}
(u_s)_{iI} = 
e^{-\frac{1}{2} \sigma_j} z^{l-1} \left( l - \frac{1}{2} r \p_r \sigma_j \right) \delta_{ij} \delta_{IJ},
\eeq
where $r=|z|$. 
As we have seen above, this bosonic mode has eigenfrequency 
related to that of the fermionic mode as \eqref{eq:b_to_f}
\beq
\epsilon_b = 
 ( l - l_j )(\omega - \tilde \omega) - \mu_J + \mu_j - \omega - \tilde \omega. 
\eeq
The towers of eigenmodes \eqref{eq:LLL_b} and \eqref{eq:LLL_f}
can be interpreted as the lowest Landau levels 
in the bosonic and fermionic sectors, respectively. 
As was done in the non-linear Schr\"odinger system 
\cite{Biasi:2017pdp}, 
it would be interesting to discuss 
the low energy dynamics of such degrees of freedom
with physically distinctive properties.

\section{Summary and Discussion}
\label{sec:summary}
In this paper, we discussed 
the supersymmetric Jackiw-Pi model in the harmonic trap. 
The super-Schr\"odinger symmetry of 
the original SUSY Jackiw-Pi model is modified 
in the presence of the external background fields 
which correspond to the generalized chemical potential terms 
including the harmonic potential. 
We have seen that the 1/3 BPS states of Jackiw-Pi vortices, 
which preserve a part of the modified supersymmetry, 
are stationary configurations rotating around the origin.  
They become static when the moduli matrix is 
at the fixed points of the spacial rotation and the flavor symmetry. 
We have investigated fluctuations around the static BPS backgrounds 
and revealed the structure of supermultiplets of eigenmodes. 
In addition to the bosonic massive NG modes, 
we identified the fermionic massive NG modes associated 
with the broken modified superconformal symmetry. 
We have also found the eigenmode spectra of 
the infinite towers of supermultiplets 
corresponding to the bosonic and fermionic lowest Landau levels. 

While we have discussed one of the simplest examples of 
(modified) non-relativistic supersymmetry in the Jackiw-Pi model, 
it has been known that there exist Chern-Simons matter systems 
with extended non-relativistic supersymmetries \cite{Nakayama:2008qz,Nakayama:2008td,Nakayama:2009cz}. 
It would be interesting to investigate 
bosonic and fermionic massive NG modes 
in the extended models such as the non-relativistic ABJM model. 
Another direction to be explored is 
to clarify the relation between 
the quantum states of the Jackiw-Pi vortices 
in the harmonic potential 
and the spectrum of the chiral primary operators 
\cite{Nakayama:2008qm, Lee:2009mm}
from the viewpoint of the non-relativistic version of 
the sate-operator mapping \cite{Nishida:2007pj}. 
If we set $\mu_f + \omega + \tilde \omega = 0$ in our model, 
the explicit time dependence of 
the supersymmetry preserved by the 1/3 BPS states
disappears and hence we can compactify 
the time direction without breaking the supersymmetry.
Such a situation is quite similar to the $\Omega$-background 
\cite{Nekrasov:2002qd, Nekrasov:2003rj}
and it would be possible 
to compute certain types of superconformal indices
by using the supersymmetric localization method \cite{Pestun:2016zxk}. 
As in the case of the vortex partition functions 
in 2d $\mathcal N=(2,2)$ theories 
\cite{Fujimori:2012ab, Fujimori:2015zaa}, 
the moduli matrix method, 
which was used to describe the BPS vortex solution, 
would play a crucial role in the localization computation
and hence it is an important future work to 
investigate the structure of the space of the BPS solutions
from the viewpoint of the moduli matrix formalism 
and its relation to the ADHM like construction discussed in Ref.\,\cite{Doroud:2015fsz}.  

\section*{Acknowledgement}
This work is supported by the Ministry of Education,
Culture, Sports, Science (MEXT)-Supported Program for the Strategic
Research Foundation at Private Universities ``Topological Science''
(Grant No.~S1511006).
The work of M.~N.~is also supported in part by a Grant-in-Aid for
Scientific Research on Innovative Areas ``Topological Materials
Science'' (KAKENHI Grant No.~15H05855) from the MEXT of Japan, and by the Japan Society for the Promotion of Science
(JSPS) Grant-in-Aid for Scientific Research (KAKENHI Grant
No.~16H03984). We would like to thank 
Yunguo Jiang and Sven Bjarke Gudnason 
for helpful discussions at an early stage of this work. 


\end{document}